# Image Authentication Based on Neural Networks

Shiguo Lian

SAMI Lab, France Telecom R&D Bejing
Beijing, P.R China, 100080
shiguo.lian@orange-ftgroup.com

**Abstract**

Neural network has been attracting more and more researchers since the past decades. The properties, such as parameter sensitivity, random similarity, learning ability, etc., make it suitable for information protection, such as data encryption, data authentication, intrusion detection, etc. In this paper, by investigating neural networks' properties, the low-cost authentication method based on neural networks is proposed and used to authenticate images or videos. The authentication method can detect whether the images or videos are modified maliciously. Firstly, this chapter introduces neural networks' properties, such as parameter sensitivity, random similarity, diffusion property, confusion property, one-way property, etc. Secondly, the chapter gives an introduction to neural network based protection methods. Thirdly, an image or video authentication scheme based on neural networks is presented, and its performances, including security, robustness and efficiency, are analyzed. Finally, conclusions are drawn, and some open issues in this field are presented.

**1 Introduction**

Neural network [1] is used to refer to artificial neural network [2], while has been used to refer to biological neural network [3]. The biological neural network is a network of biological neurons, which is in relation with nervous system. The artificial neural network is composed of artificial neurons, which is the simulation of biological neural network.

According to the learning ability, artificial neural network has been used in artificial intelligence [4]. For example, it is used to guide the robot to play chess, it is used in pattern recognition [5,6], such as pattern classification or object recognition, it is used in function approximation [7], such as time series prediction or modeling, and it is also used in data processing [8,9], such as filtering, clustering, blind signal separation and compression.

Additionally, neural network can be used to model non-linear statistical data, which can model complex relationships between inputs and outputs. Sometimes, it can generate chaos phenomenon [10]. According to this property, it has complex dynamic action, which can be used to protect data content. For example, the random sequence produced by neural network can be used to encrypt data [11], and the neural networks that generate chaos phenomenon can be used in secret communication [12].

In this chapter, some of neural network's properties suitable for content protection are introduced, some content protection schemes based on neural network is analyzed, a multimedia authentication scheme based on neural network is proposed together with some evaluation on its performances, and future work is also presented.

The rest of the paper is arranged as follows. In Section 2, some properties of neural network are presented. Some content protection schemes based on these properties are given in Section 3. In Section 4, the multimedia authentication scheme based on neural network is proposed. And its performances are analyzed in Section 5. In Section 6, some open issues in neural network based content protection are presented. And the conclusions are drawn in Section 7.

## 2 Neural Network's Properties Suitable for Content Protection

### 2.1 Learning Ability

Neural network has the possibility of learning. Given a specific task to solve and a class of functions, neural network can use a set of observations to solve the task in an optimal sense. Generally, according to the learning task, neural network's learning ability can be classified into three categories, i.e., supervised learning, unsupervised learning and reinforcement learning [13][14][15]. Supervised learning is the learning with a "teacher" in the form of a function that provides continuous feedback on the quality of solutions. These tasks include pattern classification, function approximation and speech or gesture recognition, etc. Unsupervised learning refers to the learning with old knowledge as the prediction reference. These tasks include estimation problem, clustering, compression or filtering. Reinforcement learning refers to the learning with dynamic estimation and decision. These tasks include control tasks, games and other decision making tasks. This property can be used to detect the intruders that enter a secret system without permission.

### 2.2 One-way Property

One-way property means that it is easy to compute the output from the input while difficult to compute the input from the output. In neural network, there are often more inputs than outputs, which make it difficult to decide the inputs when knowing only the outputs. Taking a simple neuron model for example, the input is composed of n elements $p_0, p_1, \ldots, p_{n-1}$, while the output is a unique element c. It is defined as

$$c = f(\sum_{j=0}^{n-1} w_j p_j + b ). \qquad (1)$$

As can be seen, it is easy to compute c from $P=[p_0, p_1, \ldots, p_{n-1}]$, while difficult to compute P from c. The difficulty is equal to solve a singular equation. Thus, it is a one-way process from the input P to the output c. This property is often required by hash functions [16][17] that are used to authenticate data's integrity.

### 2.3 Random Similarity

According to the complex relation between the nods of neural network, neural network can produce the sequences with random properties. For example, some neural networks have chaotic dynamics [10,18,19], which produce the output sequence with random properties [20], as shown in Fig. 1. Here, the sequence's autocorrelation is similar to impulse function, while the cross-correlation is near to zero.

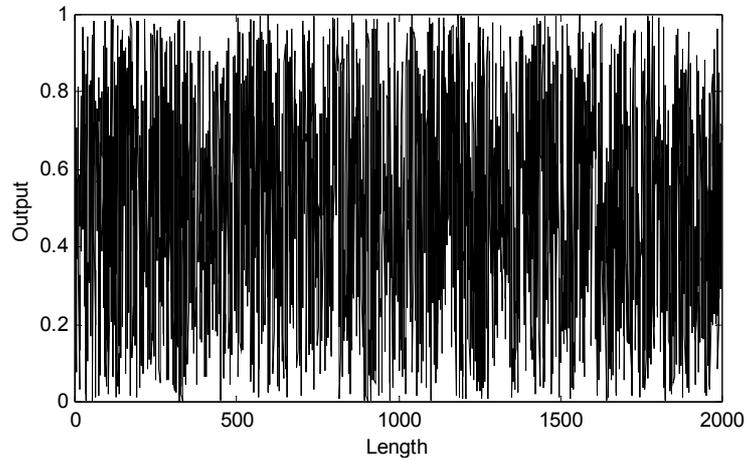

**Fig. 1** The sequence produced by a chaotic neural network

**2.4 Parameter Sensitivity**

For the neural network that has chaotic dynamics, its output is often sensitive to the inputs or such control parameter as weight [18]. It is caused by the parameter sensitivity of chaos system [21]. Taking the neural network proposed in [10] for example, although there is a slight difference in the initial value, the output changes greatly after more than 20 times of iteration, as shown in Fig. 2. This property makes the initial value suitable for the key that controls the data encryption or decryption.

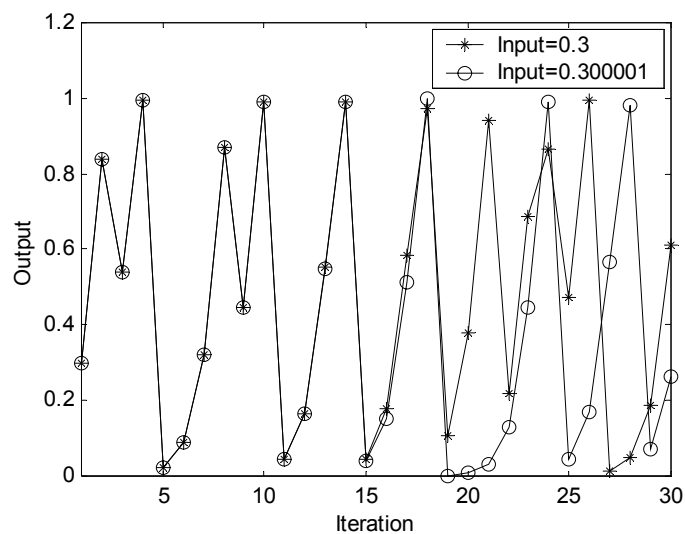

**Fig. 2** The initial-value sensitivity of a chaotic neural network

# 3 Content Protection Based on Neural Networks

## 3.1 Intrusion Detection

According to the learning ability, neural network has been used in intrusion detection [22]. Intrusion detection is an important technology in network security, which can detect illegal intruders or illegal intrusions. Using neural network's supervised learning, as shown in Fig. 3, the intrusive operations can be distinguished from normal operations [23]. Furthermore, if some classification information is provided before hand, the intrusive operations can even be classified [24]. The unsupervised learning and reinforcement learning can be used to detect new intrusive operations automatically.

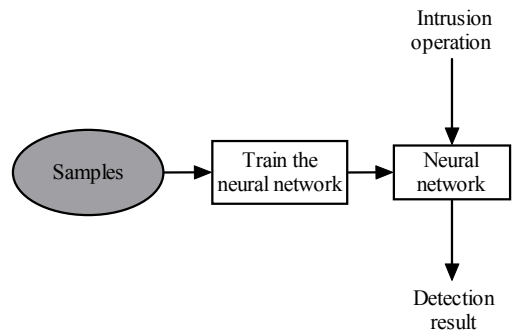

**Fig. 3** Intrusion detection based on neural network

## 3.2 Data Authentication

The one-way property makes neural network a suitable choice for hash function [16][17]. Hash function is a technique for data integrity authentication, which takes a message of arbitrary length as input and produces an output of fixed length. The hash value is often much shorter than the message, which makes it suitable for digital signature or data authentication. As a hash function, it should be easy to compute the hash value from the message, while difficult to compute the message from the hash value. This property is called one-way property. According to this case, neural network may be used to design hash function. Till now, some hash functions based on neural network have been presented [25][26], which were reported to have some advantages compared with existing schemes, such as high time-efficiency or flexible extension. For example, in [25], a hash function based on multi-layer neural network is presented, as shown in Fig. 4, which produces the hash value with the message as input.

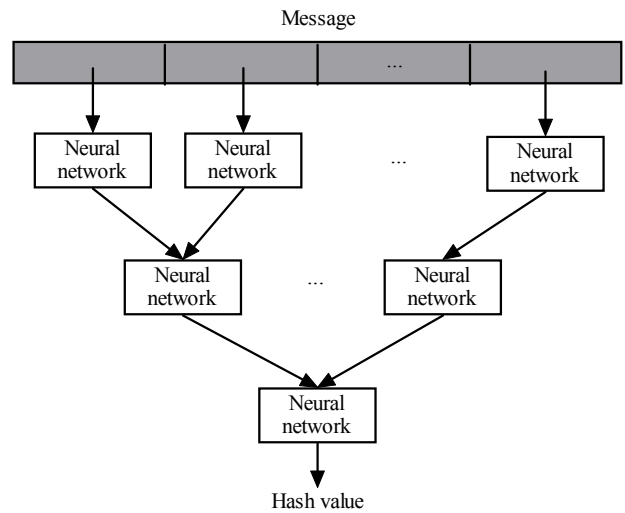

**Fig. 4** A hash function based on neural network

**3.3 Data Encryption**

According to the properties of random similarity and parameter sensitivity, neural network has been used in data encryption [20]. Data encryption is the technology that converts a message into unintelligible form under the control of the key. Till now, two kinds of encryption algorithms have been reported, i.e., stream cipher and block cipher. Stream cipher uses neural network to generate a pseudorandom sequence that modulates a message and uses the parameter as the key. The general architecture is shown in Fig. 5. For example, the ciphers [27,28,29] are constructed based on the random sequences generated from the neural networks. As a combination of neural networks and chaos, chaotic neural networks (CNN) are used in secret communication [30]. Chan et al. [31] took advantage of Hopfield neural networks' chaotic property to produce pseudo-random sequence that is used to encrypt messages, Karras et al. [32] proposed the method to evaluate the property of the pseudorandom sequence generated from chaotic neural networks, and Caponetto [33] designed a secure communication scheme based on cell neural networks. Block cipher makes use of neural networks' properties to encrypt a message block by block. In this cipher, the initial value or control parameter acts as the key. For example, Yue et al. [34] constructed the image encryption algorithm based on chaotic neural networks, Yen et al. [35] used perception neurons to construct a block cipher, Lian et al. [36] proposed an image cipher based on chaotic neural network, and Yee et al. [37] designed a symmetric block cipher based on the multi-layer perception network.

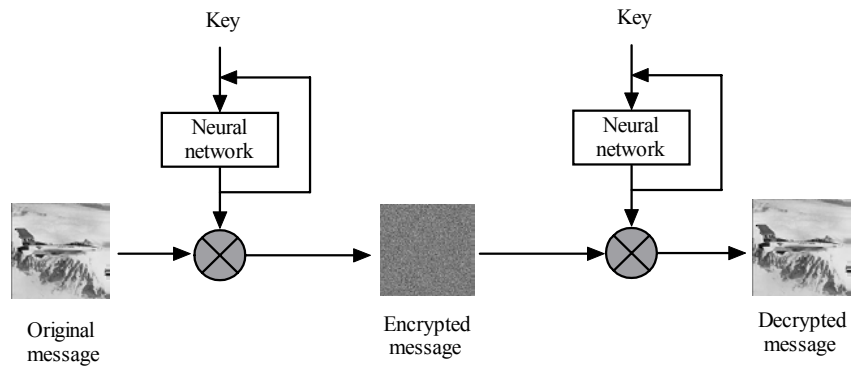

**Fig. 5** Architecture of the stream cipher based on neural network

## 4 Multimedia Content Authentication Based on Neural Networks

With the development of multimedia technology and network technology, multimedia data are used more and more widely, such as digital camera, video conference, video-on-demand, audio broadcasting, etc. These applications are in relation with human's activities in commerce, life, education, politics or even military. Thus, the security of multimedia content becomes urgent. For example, an image may be distributed from one person to another without control over Internet. During the distribution, the image's content may be tampered maliciously. For example, the person in an image is replaced by another one, an object in the image is removed, or some new objects are added to the image. To tell whether the image is tampered or not, multimedia content authentication technique [38,39] is invented, which generates authentication code from the original image, compare it with the new code generated from the received image, and tell the authentication result.

Based on the one-way property and learning ability of neural network, the multimedia content authentication scheme based on neural network is presented. In the following content, the scheme is described in detail, and its performances are also evaluated.

### 4.1 Architecture of the Proposed Scheme

The proposed multimedia content authentication scheme is shown in Fig. 6. Here, the media data, original authentication code and key are used to feed a neural network, which produces a secret parameter. Compared with media data, the secret parameter is of small size. Then, the secret parameter and the key are stored or transmitted in a secure way, while the media data are distributed freely. During distribution, media data may be tampered maliciously.

In authentication, the received media data, secret parameter and key are used to feed the same neural network, which produces the computed authentication code. By comparing the original authentication code and the computed one, the authentication result is produced. That is, if there is only slight difference between them, then the multimedia data are not tampered, otherwise, they are tampered. To authenticate multimedia data successfully, two conditions are required. Firstly, the secret parameter and key are correct. Secondly, the received media data are same to or not very different from the original media data.

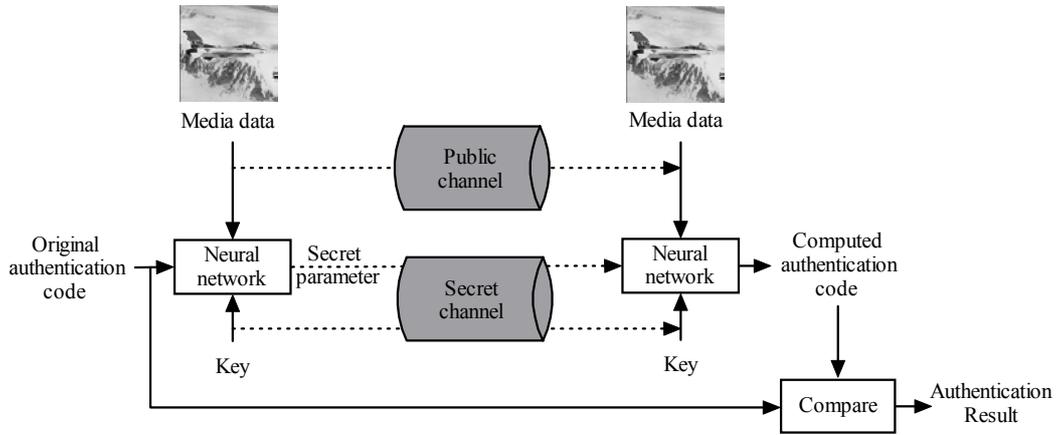

**Fig. 6** Architecture of the proposed multimedia content authentication scheme

**4.2 The Scheme Based on a Simple Neuron**

Based on a simple neural network, a practical content authentication scheme is proposed, as shown in Fig. 7(a). In this scheme, media data are partitioned into blocks, and each bit of the authentication code corresponds to a block and a simple neuron. For each simple neuron, a code bit s (s=0 or 1) and the media block composed of n pixels act as the input and control parameter. The pixels of a media block are normalized into the pixels $p_0$, $p_1$, …, $p_{n-1}$ ranging in [-1,1] by subtracting the mean and dividing the maximal amplitude. The random sequence $w_0$, $w_1$, …, $w_{n-1}$ ($w_i \in [0,1]$, i=0,1,…,n-1) is generated by a random number generator with K as the random seed. The received media block is $p'_0$, $p'_1$, …, $p'_{n-1}$, and the detected watermark bit is s'. The simple neuron shown in Fig. 7(b) is denoted by

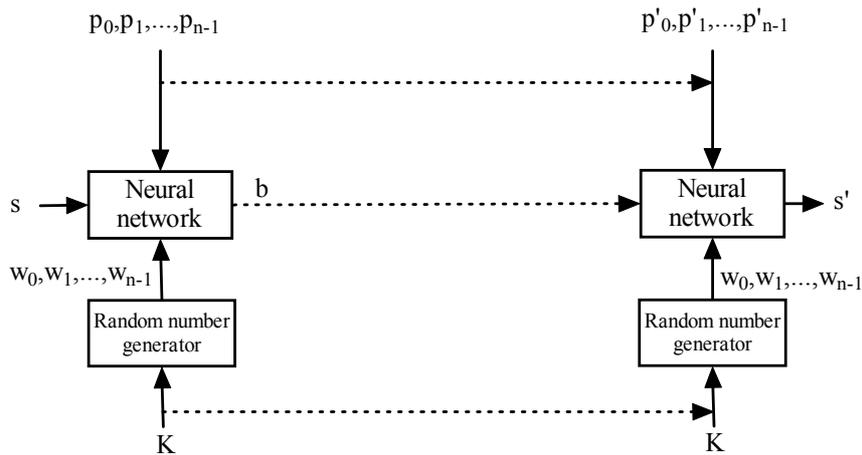

(a) The practical authentication scheme based on neural network

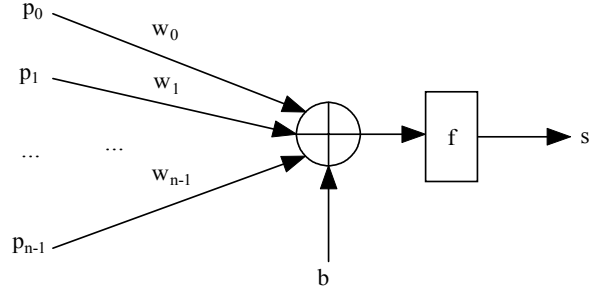

(b) A simple neuron

**Fig. 7** The proposed practical scheme based on neural network

$$s = f(\sum_{j=0}^{n-1} w_j p_j + b) = Sign(\sum_{j=0}^{n-1} w_j p_j + b) = \begin{cases} 1, & \sum_{j=0}^{n-1} w_j p_j + b > 0 \\ 0, & \sum_{j=0}^{n-1} w_j p_j + b \leq 0 \end{cases}. \quad (2)$$

**4.3 Generation of Secret Parameter**

To compute the secret parameter, the neuron layer is fed by the parameters $p_0, p_1, \ldots, p_{n-1}, w_0, w_1, \ldots, w_{n-1}$ and s. That is, by solving Eq. (2), we get

$$b = \begin{cases} > -\sum_{j=0}^{n-1} w_j p_j, & s = 1 \\ \leq -\sum_{j=0}^{n-1} w_j p_j, & s = 0 \end{cases}. \quad (3)$$

Set T as an adjustable parameter ranging in [0, 0.5]. Then, we get

$$b = \begin{cases} T - \sum_{j=0}^{n-1} w_j p_j, & s = 1 \\ -T - \sum_{j=0}^{n-1} w_j p_j, & s = 0 \end{cases}. \quad (4)$$

Here, T can be fixed or determined by experiments. For T is in relation with the robustness of the authentication system, it will be discussed in detail in Section 5.

**4.4 Content Authentication**

As shown in Fig. 7(a), the new authentication code is computed according to the following method.

$$s' = f(\sum_{j=0}^{n-1} w_j p'_j + b) = Sign(\sum_{j=0}^{n-1} w_j p'_j + b) = \begin{cases} 1, & \sum_{j=0}^{n-1} w_j p'_j + b > 0 \\ 0, & \sum_{j=0}^{n-1} w_j p'_j + b \leq 0 \end{cases}. \quad (5)$$

Here, the neuron layer is fed by the parameters $p'_0$, $p'_1$, …, $p'_{n-1}$, $w_0$, $w_1$, …, $w_{n-1}$ and b. That is, to get s' from Eq. (5). It is straightforward.

Thus, if the media data $p_j'$ (j=0,1,…,n-1) is not changed compared with $p_j$ (j=0,1,…,n-1), then it is easy to get

$$s' = \begin{cases} 1, & T > 0 \\ 0, & -T \leq 0 \end{cases}. \quad (6)$$

Thus, the authentication code can be extracted correctly.

Otherwise, the media data are changed, and the authentication code is determined by

$$s' = \begin{cases} 1, & T + (\sum_{j=0}^{n-1} w_j p'_j - \sum_{j=0}^{n-1} w_j p_j) > 0 \\ 0, & -T + (\sum_{j=0}^{n-1} w_j p'_j - \sum_{j=0}^{n-1} w_j p_j) \leq 0 \end{cases}. \quad (7)$$

In this case, the authentication code is determined by the change. If the change is smaller than T, the authentication code can still be computed correctly. Otherwise, the wrong code will be computed.

**5 Performance Evaluation**

**5.1 Security**

In the proposed scheme, the authentication code is computed under the control of the key and the secret parameter. Without them, it is difficult to compute the authentication code correctly. Thus, the security of the proposed scheme depends on two aspects, i.e., the secrecy of the key and parameter, and the brute-force attack (when not knowing the key and the parameter).

In the first aspect, the key K and the parameter b can be transmitted over a secret channel. For example, they are encrypted with such ciphers as DES, RSA, AES, etc [20]. These ciphers keep the security of the key and the parameter.

In the second aspect, for the attackers not knowing the key and the parameter, they must guess the parameters $w_0$, $w_1$, …, $w_{n-1}$, b or solve the following equation.

$$s' = \begin{cases} 1, & \sum_{j=0}^{n-1} w_j p'_j + b > 0 \\ 0, & \sum_{j=0}^{n-1} w_j p'_j + b \leq 0 \end{cases}. \quad (8)$$

However, for not knowing these parameters, it is not easier than guess the authentication bit directly with brute-force attack. For example, if N bits are computed from an image, then the brute-force space is $2^N$. It is difficult when N is big enough, e.g., no smaller than 64.

**5.2 Sensitivity to Malicious Tampering**

The authentication scheme can detect the regions tampered by malicious operations. An example is shown in Fig. 8. Here, Fig. 8(a), (b) and (c) show the original images, tampered images and detected tampered regions, respectively. In Fig. 8(c), the black area denotes the tampered regions.

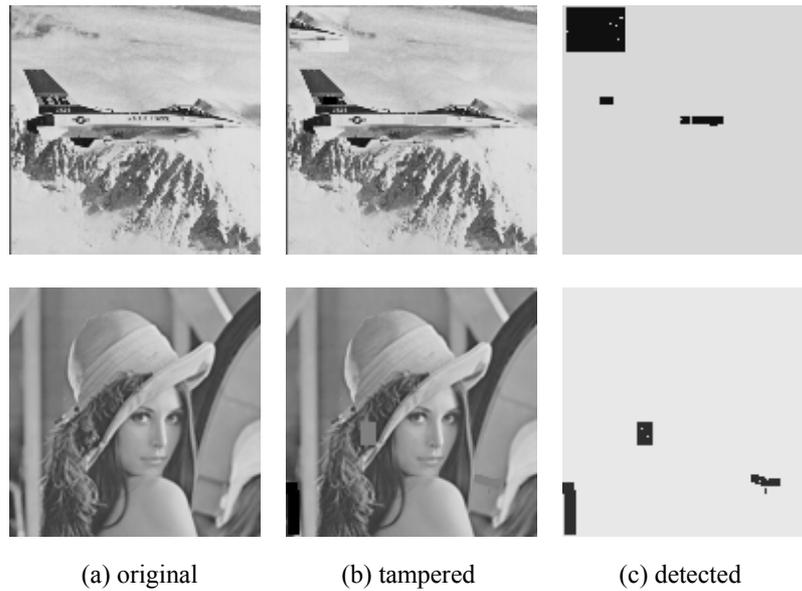

(a) original (b) tampered (c) detected

**Fig. 8** Detection of malicious tampering

**5.3 Robustness**

The authentication code can be computed correctly when media data are not changed during transmission or distribution. The proposed scheme is robust to some operations applied to the media data, such as adding noise or recompression. The performance depends on two parameters, i.e., the adjustable parameter T and the block size B.

*5.3.1 Robustness against noise*

The block size affects the correct detection rate (Cdr) when the media data are changed with some operations. Generally, the bigger the size is, the higher the correct detection rate is. Fig. 9(a) shows the relation between the block size and the correct detection rate. Here, Gaussian noise is added to the media data, B varies from 4 to 32, and T=0.2. As can be seen, for the same noise strength, the correct detection rate increases with the rise of the block size.

The adjustable parameter T also affects the correct detection rate (Cdr) in the condition of adding noise. Fig. 9(b) shows the relation between T and Cdr. Here, Gaussian noise is added, T varies from 0 to 0.3, and B=8. As can be seen, the correct detection rate increases with T when the noise strength is certain. When B≥8, T≥0.2 and Gaussian noise's variance is no bigger than 0.01, the correct detection rate is often no smaller than 80%.

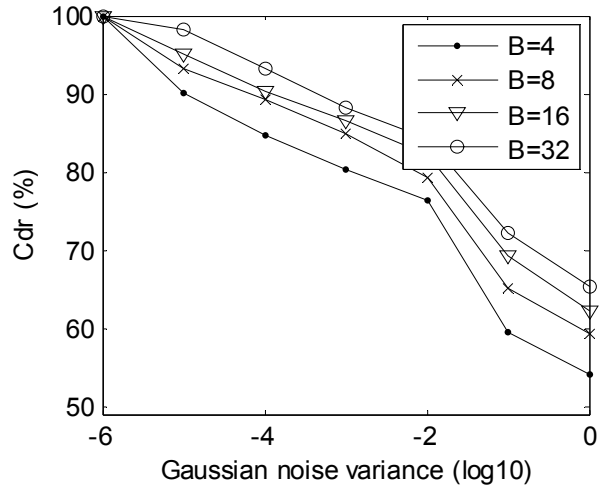

(a) Relation between Cdr and B

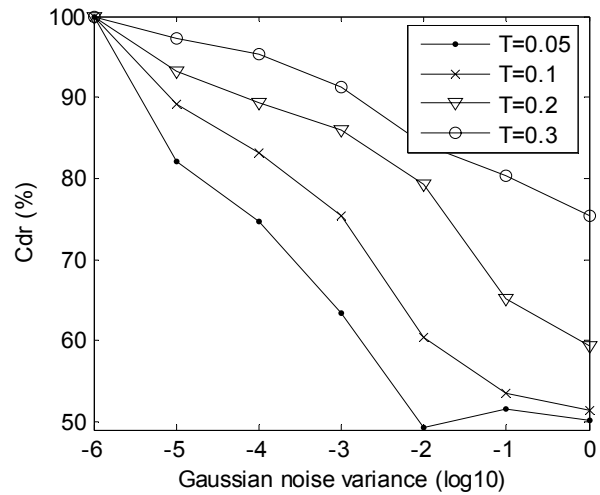

(b) Relation between Cdr and T

**Fig. 9** Robustness against adding noise

*5.3.2 Robustness against recompression*

The proposed scheme is robust against JPEG compression. The correct detection rate (Cdr) is in relation with the adjustable parameter T. Fig. 10 shows the relation between Cdr and T. Here, the image is Lena 512*512, the compression quality ranges from 100 to 40, T ranges from 0 to 0.3, and B=8. As can be seen, Cdr increases with the rise of T if the compression quality is certain. When T≥0.2 and the compression quality is no smaller than 70, the correct detection rate is no smaller than 85%.

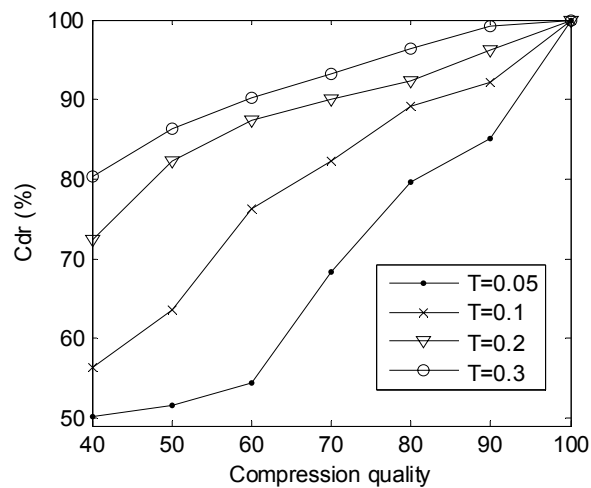

**Fig. 10** Robustness against JPEG Compression

*5.3.3 Some means to improve the robustness*

According to the experiments and analyses, the following means can be adopted to improve the correct detection rate when the media data are operated. The first one is to increase T, the second one is to enlarge the block size, and the third one is to compute the authentication bit repeatedly. It should be noted that the detection precision will be reduced by using the latter two means.

**5.4 Time Efficiency**

The computing or authentication process can be implemented efficiently. Table 1 shows the results on some images. Here, the computer is of 1.7GHz CUP and 256M RAM, and the computing/authentication program is implemented in C code. As can be seen, the computing/authentication time is no more than 1 second, when the image's size is no bigger than 512×512. Considering that each block can be computed or authenticated independently, neural networks' parallel properties can be used to improve the efficiency of the implementation.

Table 1 Efficiency of the Embedding or Detection Process

| Image | Size | Color/Gray | Computing (s) | Authentication (s) |
|---|---|---|---|---|
| Lena | 128×128 | Gray | 0.1 | 0.1 |
| Boat | 256×256 | Gray | 0.3 | 0.3 |
| Village | 512×512 | Gray | 0.6 | 0.4 |
| Lena | 128×128 | Color | 0.2 | 0.2 |
| Peppers | 256×256 | Color | 0.4 | 0.3 |
| Baboon | 512×512 | Color | 0.8 | 0.9 |

## 6 Open Issues

Using neural network to protect data content is straightforward in some extent because of neural network's properties. For example, the learning ability enables neural network to know and identify intrusion operations, and the learning ability enables neural network to understand or even remember multimedia content. However, for artificial neural network's intelligence is much lower than biological neural network, there are still some open issues in the engineering applications, especially, in content protection.

Firstly, neural network has some properties suitable for content protection, but which property is suitable for which protection scheme is still uncertain. In this chapter, only four properties are introduced, including learning ability, one-way property, random similarity and parameter sensitivity. Maybe, several properties are all required by one scheme. It needs the cooperation between both the researchers in neural network and the researchers in content protection.

Secondly, neural network's properties should be investigated before being used in content protection. Not all neural networks are suitable for content protection. Some principles need to be constructed to guide the design of neural network based content protection.

Thirdly, as a content protection scheme, the performances should be evaluated thoroughly before being used in practical applications, which includes the scheme based on neural network. Till now, some means have been proposed to evaluate the new schemes based on neural network [40]. However, the work is far from enough.

Fourthly, using the learning ability to authenticate images is still challenging. It is expected for neural network to learn and remember an image and later to tell whether the tampered copy is illegal or not. Although it is challenging, artificial neural network's development may make it true.

## 7 Conclusions and Future Work

In this paper, some existing research work in neural network based content protection is firstly introduced, including neural network's properties suitable for content protection and some content protection schemes based on neural network. Then, a multimedia content authentication scheme is proposed, which makes use of neural network's learning ability and one-way property to detect malicious tampering on multimedia content. The performance evaluation shows the scheme's practicality. Furthermore, some open issues in this research field are presented. Finally, some conclusions are drawn.